\documentclass[prl,twocolumn,showpacs,amsmath,amssymb,superscriptaddress,longbibliography]{revtex4-1}
\usepackage{psfrag,graphicx}
\usepackage{amsfonts,amssymb,amsmath}      
\usepackage{graphicx}
\usepackage{dcolumn}
\usepackage{bm}
\usepackage{float}
\usepackage{hyperref}
\usepackage{accents}
\usepackage{bbding}
\usepackage{color,soul}
\usepackage{epstopdf}
\usepackage{changepage}
\usepackage{orcidlink}
\usepackage[normalem]{ulem}
\usepackage{dsfont}

\newcommand{\erf}[1]{Eq.~(\ref{#1})}
\newcommand{\beq}{\begin{equation}}
\newcommand{\eeq}{\end{equation}}

\newcommand{\dg}{^\dagger}

\newcommand{\smallfrac}[2]{\mbox{$\frac{#1}{#2}$}}

\newcommand{\half}{\smallfrac{1}{2}}
\newcommand{\bra}[1]{\langle{#1}|}
\newcommand{\ket}[1]{|{#1}\rangle}
\newcommand{\ip}[2]{\langle{#1}|{#2}\rangle}
\newcommand{\op}[2]{\ket{#1}\bra{#2}}

\newcommand{\Tr}{\text{Tr}}

\newcommand{\tp}{^{\top}}

\newcommand{\s}[1]{\hat{\sigma}_{#1}}

\newcommand{\ex}[1]{\langle{#1}\rangle}
\newcommand{\dd}{{\rm d}}

\newcommand{\ie}{{\em i.e.}}
\newcommand{\eg}{{\em e.g.}}

\newcommand{\inv}{^{-1}}

\newcommand{\incoh}{_{\cal I}}
\newcommand{\coh}{_{\cal C}}

\renewcommand{\bar}[1]{\overline{#1}}

\definecolor{nblue}{rgb}{0.06,0.3,0.73}
\definecolor{nblack}{rgb}{0,0,0}
\definecolor{nred}{rgb}{0.9,0.1,0.1}
\definecolor{nmagenta}{rgb}{0.7,0.0,0.3}
\definecolor{applegreen}{rgb}{0.55, 0.71, 0.0}
\definecolor{nteal}{rgb}{0, 0.5, 0.5}

\newcommand{\blk}{\color{nblack}}

\newcommand\stPW{\bgroup\markoverwith{\applegreen{\rule[0.5ex]{2pt}{0.4pt}}}\ULon}

\begin{document}

\title{An Energetic Constraint for Qubit-Qubit Entanglement}
\author{Kiarn T. Laverick\orcidlink{0000-0002-3688-1159}}
\affiliation{MajuLab, CNRS-UCA-SU-NUS-NTU International Joint Research Laboratory}
\affiliation{Centre for Quantum Technologies, National University of Singapore, 117543 Singapore, Singapore}
\author{Samyak P. Prasad\orcidlink{0009-0001-3672-6001}}
\affiliation{MajuLab, CNRS-UCA-SU-NUS-NTU International Joint Research Laboratory}
\affiliation{Centre for Quantum Technologies, National University of Singapore, 117543 Singapore, Singapore}
\author{Pascale Senellart\orcidlink{0000-0002-8727-1086}}
\affiliation{Universit\'e Paris-Saclay, Centre de Nanosciences et de Nanotechnologies, CNRS, 10 Boulevard Thomas Gobert, 91120, Palaiseau, France}
\author{Maria Maffei\orcidlink{0000-0001-5183-4716}}
\affiliation{Universit\'e de Lorraine, CNRS, LPCT, F-54000 Nancy, France}
\author{Alexia Auff\`eves\orcidlink{0000-0003-4682-5684}}
\email{alexia.auffeves@cnrs.fr}
\affiliation{MajuLab, CNRS-UCA-SU-NUS-NTU International Joint Research Laboratory}
\affiliation{Centre for Quantum Technologies, National University of Singapore, 117543 Singapore, Singapore}

\date{\today}

\begin{abstract}

We uncover a fundamental energetic constraint governing qubit-qubit entanglement. By decomposing a qubit’s internal energy into coherent and incoherent components, we define a coherent-energy deficit as the loss of coherent energy relative to the pure state with the same mean energy. We show that, for any pure two-qubit state, this deficit is proportional to the tangle, establishing a mathematical equivalence between a locally measurable energetic quantity and an entanglement monotone. This equivalence reveals a new conserved quantity, the coherent-energy budget, that can be redistributed between local coherence and entanglement during local-energy-preserving dynamics. We derive tight bounds on the efficiency of local coherence-to-entanglement conversion and on the reverse extraction of local coherence from entanglement in these processes, and show how mixedness impacts these bounds. Extending the framework to three-qubit systems shows how local coherent energy can be successively converted into bipartite and tripartite entanglement under local-energy-preserving dynamics. These results identify a conserved energetic quantity governing coherence-entanglement interconversion and provide a quantitative description of the associated constraints.

\end{abstract}

\maketitle

{\em Introduction.}---Entanglement is a key resource for quantum technologies. The question of its fundamental resource cost has historically been explored via entropic metrics \cite{Bennett96,HillWootters97,Amico08,HorodeckiRMP09,NieChu10}, leading to thermodynamical understandings, such as second laws for entanglement \cite{Horodecki02,Ganardi25}. More recently, the focus has been turned to energy costs, which have been explored from multiple perspectives, from entanglement distillation~\cite{Das2017canonical} and  generation~\cite{Galve2009,Navarrete12,Huber2015,Bruschi15,Piccione20}, possibly under energy constraint~\cite{Chiribella2017, deOliveira24}, to its extraction~\cite{beny2018energy,hackl2019minimal}, and distribution~\cite{horodecki2025quantification}, with potential application to the energetics of quantum networks \cite{Yehia24}. In this paper, we bring a new conceptual tool to these research lines: we introduce a structure to qubits' internal energies and explore the impact of qubit-qubit entanglement on this structure. 

A qubit's internal energy is proportional to the population of its excited state. It can always be analyzed in terms of a coherent and an incoherent part, the former (the latter) stemming from the average ``dipole'' of the qubit's transition (the dipole fluctuations)~\cite{Cohen2024atom}. The coherent energy can be extracted, by mapping the qubit's state onto the zero and one-photon Fock states of an initially empty bosonic mode, that plays the role of an uncharged battery~\cite{monsel2020energetic}. This can be done, \eg~through the spontaneous emission of the qubit in an initially empty waveguide. Then, the coherent energy is accessible as the energy locked in the coherent component of the emitted field~\cite{monsel2020energetic,wenniger2023}. This makes it directly measurable through homodyne and heterodyne measurements (further referred to as ``-dyne measurements")~\cite{Carmichael93}. 


When a qubit is in a mixed state, or entangled with another one, its coherent energy is lower than that of the reference pure state with the same mean energy. We show that, for any pure two-qubit state, the resulting coherent-energy deficit of each qubit is exactly proportional to the squared concurrence, or tangle \cite{CofWoo00}, establishing a direct correspondence between a locally measurable energetic quantity and an entanglement monotone. Equivalently, the coherent-energy deficit and the coherent energy of each qubit add up to the coherent energy of the corresponding reference pure state, which we refer to as the coherent-energy budget. For local-energy-preserving (LEP) processes, \ie~processes that conserve the {\em mean} energy of each qubit, this budget is conserved. This conservation law provides a direct energetic expression of the trade-off between local coherence (\ie, the coherence in each qubit) and entanglement \cite{Andolina2019,Shi2022Entanglement,Horo2002,Toni2015}. It yields tight bounds on the amount of entanglement that can be generated from local coherence and, conversely, on the amount of local coherence that can be recovered from entanglement in LEP processes. We characterize the associated conversion efficiencies and illustrate them with a simple controlled-phase gate. Finally, we show how mixedness reduces these efficiencies by limiting the reversible conversion between coherence and entanglement.

We finally move to a configuration where the state of a qubit is successively mapped onto two other qubits through LEP two-qubit interactions. Information is thus progressively redundantly encoded onto non-local logical qubits made of two, then three entangled physical qubits. Our framework reveals a similar energetic trade-off between two-qubit and three-qubit entanglement, the whole sequence appearing as a transfer of coherent energy from local qubits to increasingly non-local logical qubits. \\

\begin{figure}
   \centering
  \includegraphics[width=1.05\linewidth]{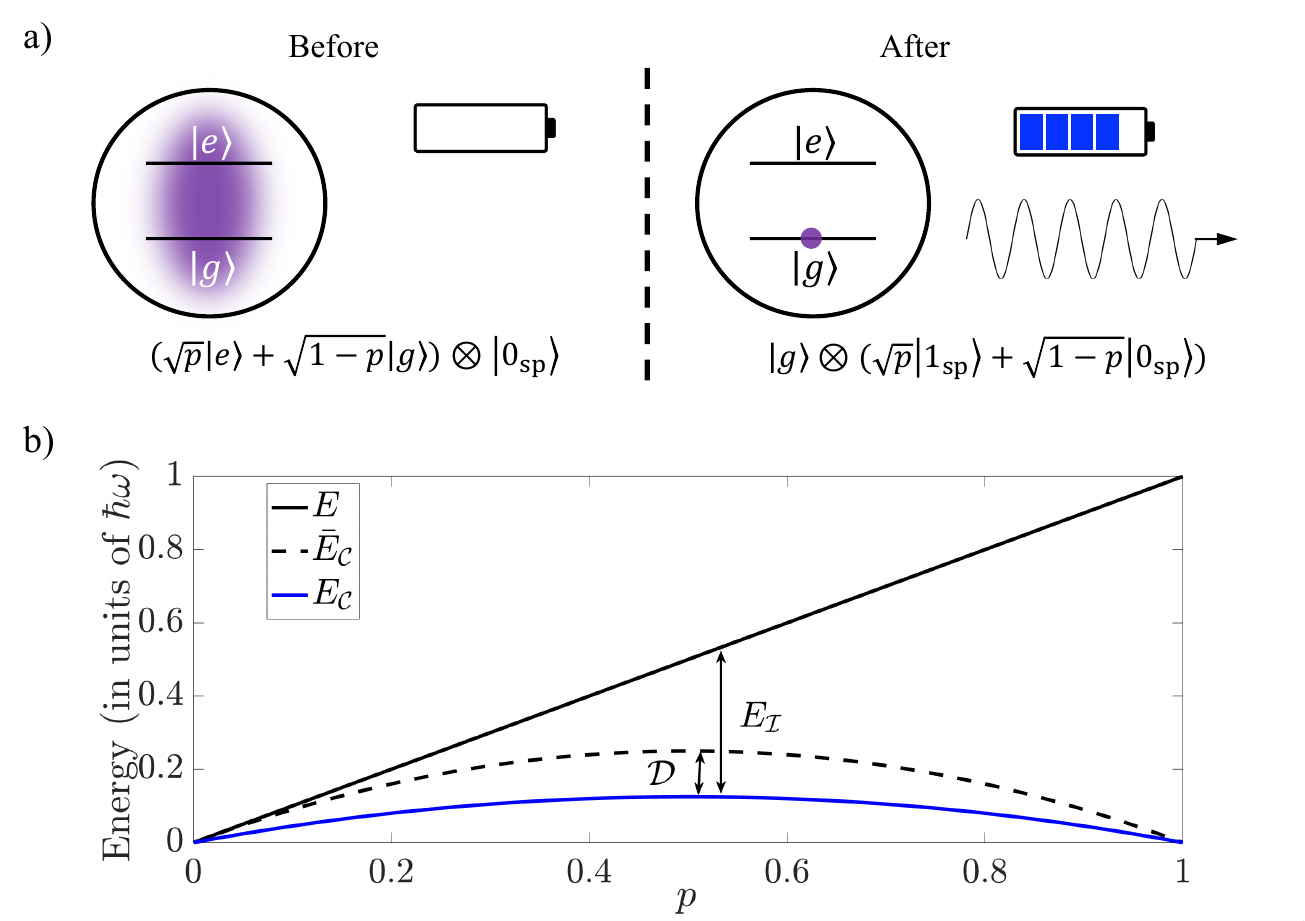}
    \caption{a) An illustration depicting a method for extracting a qubit's coherent energy. The qubit's state is mapped onto the state of a bosonic mode $a_{\rm sp}$, which plays the role of an initially uncharged battery. Doing so, the coherent energy corresponds to the energy locked in the coherent component of the field, which can be subsequently measured through -dyne type measurements. b) The energetic splitting into coherent $E\coh$ and incoherent $E\incoh$ components for a qubit with mean energy $E=\hbar \omega p$. The coherent energy  (blue-solid line) is lower than the coherent-energy budget $\bar{E}\coh$ (black-dashed line) due to decoherence (see main text). The difference between the two lines is the coherent-energy deficit ${\cal D}$. Here we have fixed $E\coh = \bar{E}\coh/2$. }
   \label{fig:Eg_Energetics}
\end{figure}

\noindent {\em Energetics of a single qubit.}---Let us consider a qubit of excited (ground) state $\ket{e}$ ($\ket{g}$), dipole operator $\s{}=\ket{g}\bra{e}$ and Hamiltonian $\hat{H} = \hbar \omega \s{}^\dagger \s{}$, where $\omega >0$ denotes the transition frequency. A general qubit state is described by a density matrix $\rho=\sum_{i,j \in\{g,e\}}\rho_{ij}\ket{i}\bra{j}$, where the population of the excited state is denoted $\rho_{ee} = p$. The mean qubit energy reads $E=\hbar \omega p$. We split this energy into two components, $E = E\coh + E\incoh$, with $E\coh \equiv \hbar \omega |\rho_{eg}|^2= \hbar \omega |\langle \s{} \rangle|^2$. We refer to $E\coh$ as the {\it coherent energy} owing to its direct relation with the coherence of the qubit's density matrix. Introducing $\delta \s{} = \s{} - \langle \s{} \rangle$, the incoherent energy reads $E\incoh = \hbar \omega \langle \delta \s{}^\dagger \delta \s{} \rangle$ and quantifies the remnant energy locked in the dipole's fluctuations. 

The coherent and total energies can be deduced from a simple qubit tomography. As detailed in the End Matter (EM), $E\coh$ is of a fundamentally different nature than the ergotropy, which is the maximal amount of work extractable through unitary operations \cite{Binder2018}. Indeed, unitary operations rely on a classical drive, which can stem from an initially loaded battery \cite{monsel2020energetic}. On the opposite, the coherent energy can be extracted by letting the qubit spontaneously release its energy in an initially empty waveguide, that plays the role of an unloaded battery (See Fig.\ref{fig:Eg_Energetics}a). Along the process, the qubit's coherent energy is transferred to the energy contained in the coherent component of the emitted field, such that $E\coh= \hbar \omega |\langle a_\text{sp} \rangle|^2$. $a_\text{sp}$ refers to the temporal mode of the spontaneous wavepacket (See EM). Once extracted into this mode, the coherent energy can be measured with -dyne setups, \ie, by interfering the quantum field with a coherent field \cite{Carmichael93}. Note that this operation can be also seen as the discharge of the battery, that extracts the desired coherent energy in a load. These energy transfers have been experimentally characterized using a semiconductor quantum dot \cite{wenniger2023}.

When the qubit undergoes decoherence, its coherent energy falls below that of a reference pure state with the same mean energy $E$, further denoted $\bar{E}\coh=\hbar\omega p(1-p)$. The resulting coherent-energy deficit verifies
\beq\label{eq:wp_Energy} {\cal D} \equiv \bar{E}\coh - E\coh = \hbar \omega \det[\rho]. \eeq By construction, ${\cal D}\geq 0$, with equality for pure states only. This decomposition is illustrated in Fig.~\ref{fig:Eg_Energetics}b, where the different energy contributions are plotted as a function of $p$. The quantity $\bar{E}\coh$ is symmetric about $p=1/2$, corresponding to the point of maximal qubit coherence. $\bar{E}\coh < E$ because of fundamental phase fluctuations affecting the qubit's dipole. We used arbitrary values of $|\rho_{eg}|$ to showcase the impact of decoherence, $E\coh < \bar{E}\coh$. Complete decoherence $|\rho_{eg}|=0$ is manifested by a maximal coherent-energy deficit, ${\cal D}_\text{max}=\bar{E}\coh$.\\

\noindent {\em Two qubits: Coherence vs. entanglement trade-off.}---
We now consider two qubits $A$ and $B$ of respective frequencies $\omega_A$ and $\omega_B$, whose entanglement is quantified by Wootters' concurrence \cite{HillWootters97,Wootters98}. Unless stated otherwise, we focus on pure bipartite states $\rho_{AB}=\ket{\Psi_{AB}}\bra{\Psi_{AB}}$. In this case, the concurrence reads \cite{HorodeckiRMP09} ${\mathbb C}[\rho_{AB}] = \sqrt{2(1 - \Tr[(\rho_m)^2])}= 2\sqrt{{\rm det}\rho_m}$, where $\rho_{m}$ is the reduced state of qubit $m=A,B$. Introducing the scaled square concurrence, or {\it tangle} \cite{CofWoo00}, as $C^2={\mathbb C}^2/4$, and dropping the argument for notational simplicity, we find that the coherent-energy deficit experienced by each qubit reads 
\beq \label{Deficit-Coherence} {\cal D}^{(m)} = \hbar \omega_m \text{det}{\rho_m}=\hbar \omega_m C^2. 
\eeq 
Equation~\eqref{Deficit-Coherence} is the first result of this work. It establishes a mathematical equivalence between an energetic quantity, the coherent-energy deficit ${\cal D}^{(m)}$, and a genuine entanglement monotone, $\hbar \omega_m C^2$. From an operational perspective, Eq.~\eqref{Deficit-Coherence} implies that the entanglement shared by two qubits can be inferred from purely local energetic measurements, namely the mean and coherent energies of a single qubit, using \eg, the setup depicted in Fig.~\ref{fig:Eg_Energetics}a. Throughout this paper, ``local" shall refer to properties or operations associated with a single qubit, ``non-local" to those involving multiple qubits. More fundamentally, Eq.~\eqref{Deficit-Coherence} can be recast into an energetic constraint linking local coherence and entanglement, 
\beq \label{Deficit-Coherence2} \bar{E}\coh^{(m)} = E\coh^{(m)} + \hbar \omega_m C^2. 
\eeq 
Trade-offs between local coherence and entanglement are well-known from an entropic perspective: increasing qubit-qubit entanglement raises the von Neumann entropy of the reduced states and thereby diminishes their coherence in any basis. Yet, a direct energetic counterpart has remained elusive. Previous works identified entropy-induced reductions of ergotropy and free energy \cite{Andolina2019,Shi2022Entanglement,Horo2002,Toni2015}, but these effects are only indirectly related to quantum coherence. By contrast, Eq. (3) establishes a direct energy balance in which the local coherent energy $E^{(m)}\coh$ and the entanglement contribution $\hbar\omega_m C^2$ appear as two complementary manifestations of the same underlying resource, $\bar{E}\coh^{(m)}$. For this reason, we refer to $\bar{E}\coh^{(m)}$ as the {\it coherent-energy budget} allocated to qubit $m$.

Finally turning to the two-qubit state, we define the {\it entanglement energy} as $E_\text{ent}=(\hbar\omega_A + \hbar \omega_B) C^2$. Since $C$ is symmetric under qubit exchange, the qubit with the largest transition frequency necessarily contributes more energy to the generation of entanglement. Introducing the total local coherent energy $E\coh = E\coh^{(A)}+E\coh^{(B)}$ and the total coherent-energy budget $\bar{E}\coh = \bar{E}\coh^{(A)}+\bar{E}\coh^{(B)}$, we get the global energetic constraint which reflects at the two-qubit level the local coherence-entanglement trade-off,
\beq \label{eq:Deficit-Coherence3} 
\bar{E}\coh = E\coh + E_\text{ent}.
\eeq 
Any bipartite energetic configuration is thus characterized by its entanglement ratio $\eta_\text{ent} = E_\text{ent}/\bar{E}\coh$, with $\eta_\text{ent} \leq 1$ by construction.\\

\noindent {\em Local-energy-preserving processes.}---For each qubit, the coherent-energy budget $\bar{E}^{(m)}\coh$ is fully determined by the mean energy $E^{(m)}=\hbar \omega_m p_m$, where $p_m$ denotes the excitation probability. Eqs.~\eqref{Deficit-Coherence2} and \eqref{eq:Deficit-Coherence3} then reveal energy conservation laws along local-energy-preserving (LEP) processes: local coherence can be reversibly converted into entanglement while preserving the coherent-energy budget of each qubit. Such processes are particularly appealing from a thermodynamic standpoint, as an agent switching the qubit-qubit coupling on and off pays no net energetic cost: the work invested at switch-on is exactly recovered at switch-off. An explicit example of an LEP process will be elucidated shortly. 

\begin{figure}[t] \centering \includegraphics[width=\linewidth]{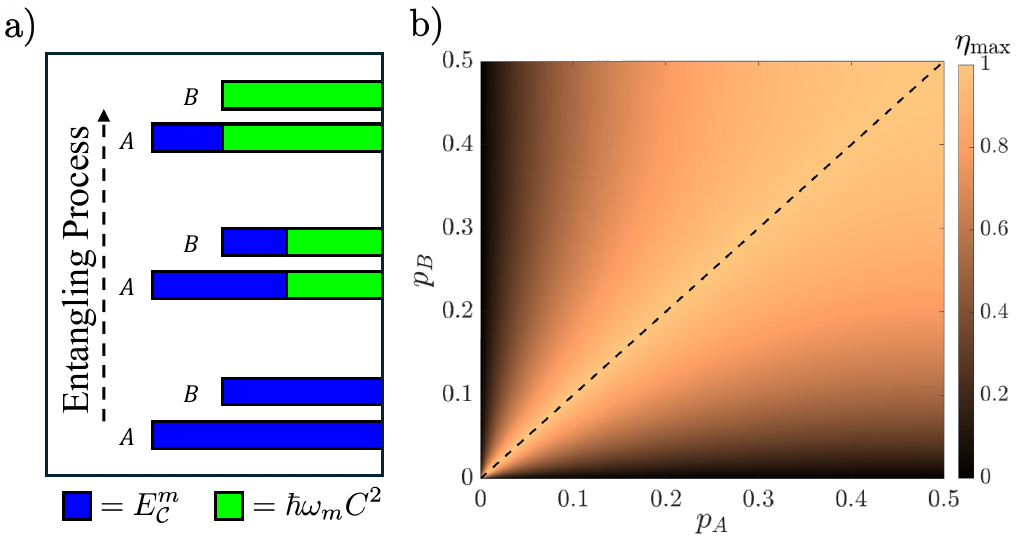} \caption{a) A diagrammatic representation of the maximal energetic conversion of coherent energy into squared concurrence or tangle. Any asymmetry in the coherent-energy resources of qubits $A$ and $B$ arising from $p_A \neq p_B$ leaves a finite amount of coherent energy trapped in one qubit. b) Maximal local-coherence-to-entanglement conversion efficiency, $\eta_{\rm max}(\{p_A,p_B\})$. Perfect conversion is achieved when the excitation is evenly distributed between the two qubits (dashed gray line). Optimal points become increasingly robust to energetic fluctuations as $p_A$ and $p_B$ approach $1/2$. Here we have taken $\omega_A = \omega_B$.} \label{fig:Eg_Efficiency} \end{figure}

In this perspective, $\eta_\text{ent}$ conveniently identifies with an efficiency of local-coherence-to-entanglement conversion for an initially factorized two-qubit state. For such a reference state, the entire coherent-energy budget is initially stored in local coherences and is therefore available for entanglement generation. The process is sketched in Fig.~\ref{fig:Eg_Efficiency}a), allowing us to derive a tight upper bound on the entanglement that can be generated from two qubits with mean energies $E^{(m)}$. Combining Eq.~\eqref{Deficit-Coherence} with the symmetry of entanglement yields \begin{equation} C^2 \leq C_\text{max}^2 = \min_{m=A,B}\{\bar{E}^{(m)}\coh / \hbar \omega_m\}. \end{equation} We choose $\bar{E}^{(B)}\coh/\hbar\omega_B \leq \bar{E}^{(A)}\coh/\hbar\omega_A$ without loss of generality. Then qubit $A$ cannot convert more than $\omega_A/\omega_B \bar{E}^{(B)}\coh$ of its coherent-energy budget into entanglement, yielding the maximal efficiency \begin{equation} \eta^{\rm max}_\text{ent} = 1- \frac{\bar{E}^{(A)}\coh-\omega_A/\omega_B\bar{E}^{(B)}\coh}{\bar{E}\coh}. \end{equation} Fig.~\ref{fig:Eg_Efficiency}b) displays $\eta^{\rm max}_\text{ent}$ for $p_A,p_B \in [0, 1/2]$ for $\omega_A = \omega_B$. Perfect local-coherence-to-entanglement conversion ($\eta^{\rm max}_\text{ent}=1$) is achieved for $p_A=p_B=p$ \cite{SM}. Up to local unitaries, the corresponding optimal states are the perfectly correlated states $\ket{\Psi_{AB}} = \sqrt{p}\ket{1_A,1_B}+\sqrt{1-p}\ket{0_A,0_B}$, with tangle $C_\text{max}^2 = p(1-p)$. Finally, maximizing over $p$ yields the Bell states at $p=1/2$. Note that the reverse process can be analyzed on equal footing: along LEP dynamics, entanglement is converted back into local coherence. In this case, the entanglement energy $E_\text{ent}$ sets the maximal amount of local coherent energy $E\coh$ that can be generated.

We have further extended the framework to mixed two-qubit states $\rho_{AB}$ (see EM). We show that ${\cal D}^{(m)} \geq \hbar \omega_m C^2$. This inequality can be recast as $\bar{E}\coh=E\coh + E_\text{ent} + {\cal L}$, where ${\cal L} \geq 0$ quantifies a loss associated with mixedness. Accordingly, mixedness lowers the amount of entanglement that can be generated from local coherence during LEP processes and reciprocally.\\

\noindent {\em Controlled-phase gate.}---We now analyze a specific example of LEP entangling process between two qubits, namely a $\pi$-phase gate. The coupling Hamiltonian reads $\hat{V}(t) = \hbar G(t) \hat{\Pi}_A \hat{\Pi}_B$, where $\hat{\Pi}_m=\s{m}\dg \s{m}$ for $m = A,B$, $G(t)=\pi /T$ for $t=[0,T]$ and $G(t)=0$ otherwise. The process is LEP since $[V(t),\sigma_z^{(m)}]=0$. It is energetically free since the qubit's mean energy increases by $\langle V(t) \rangle = \pi p_A p_B/T$ when the coupling is switched on and decreases by the same amount at the switchoff. Starting from a product state $\ket{\psi_0} = \ket{\psi_A}\otimes\ket{\psi_B}$ at $t=0$, with $\ket{\psi_m} = \sqrt{1 - p}\ket{g_m} + \sqrt{p}\ket{e_m}$, the two-qubit state at time $t$ in the interaction representation ideally reads $\ket{\psi(t)}=\sqrt{1-p}\ket{g_A}\ket{M_B^g(t)}+\sqrt{p}\ket{e_A}\ket{M_B^e(t)}$. We have introduced $\ket{M_B^g(t)} = \sqrt{1-p_B}\ket{g_B}+\sqrt{p_B}\ket{e_B}$ and $\ket{M_B^e(t)} = \sqrt{1-p}\ket{g_B}+e^{-i\pi t/T}\sqrt{p}\ket{e_B}$ the states of $B$ conditioned to the energy states of $A$. As the interaction builds correlations between the qubits, coherence in the energy basis of $A$ is progressively reduced according to $E^{(A)}\coh (t)= \bar{E}^{(A)}\coh |\epsilon^{(B)}(t)|^2$, where $\epsilon^{(B)}(t) = \langle M_B^g(t) | M_B^e(t) \rangle$ stands for the distinguishability of the $B$ states with $|\epsilon^{(B)}(t)|\leq 1$. Similar expressions can be written for $B$. Entanglement builds up until $t = T$ where $\epsilon^{(B)}(T) = 1-2p$. Thus for such process, the only way to reach a complete conversion ($E\coh^{(m)}(T)=0$) requires an initial state with $p=1/2$, yielding a Bell state at $t = T$. Note that this feature is specific to the $\pi$-phase gate and should not be regarded as a generic property of LEP processes.

\begin{figure}[t]
 \centering
    \includegraphics[width=\linewidth]{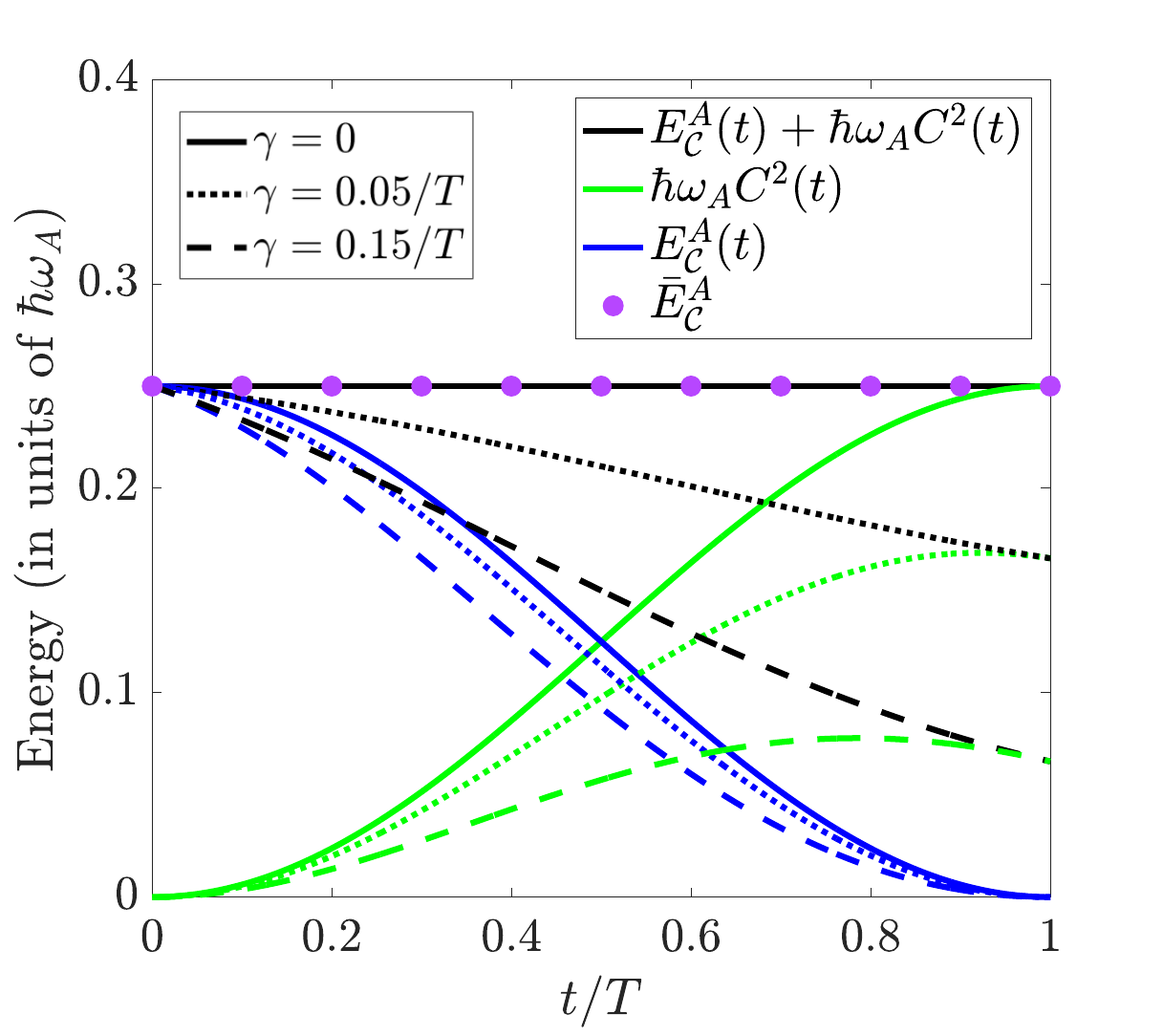}
    \caption{The energetics of a $\pi$-phase gate for three possible pure dephasing rates, $\gamma = 0$ (solid lines), $\gamma = 0.05/T$ (dotted lines) and $\gamma = 0.15/T$ (dashed lines). The bipartite qubit system is initiated in the $\ket{+_A,+_B}$ state, where $\ket{+_m} = (\ket{e_m}+\ket{g_m})/\sqrt{2}$. In the $\gamma > 0$ case, the coherent-energy budget (non-solid black lines) is strictly lower than $\bar{E}\coh$ (purple markers) due to the mixing that happens during the evolution. We observe that the entanglement energetic measure $\hbar \omega_A C^2(t)$ (green) is much more sensitive to the mixing process as $\gamma$ increases, as opposed to the coherent energy (blue)}
    \label{fig:PiGate}
\end{figure}

We have simulated the experiment for an initial state where $p=1/2$. Taking into account pure dephasing on both qubits, the LEP evolution equation of the joint state reads $\dot{\rho}_{AB} = -i[\hat{H}(t),\rho_{AB}] + {\cal L}^{(m)}[\rho_{AB}]$, with $\hat{H}(t)=\hat{H}_A+\hat{H}_B+\hat{V}(t)$ and ${\cal L}^{(m)}[\rho] = \gamma (\s{z}^{(m)} \rho \s{z}^{(m)} - \rho) $. We have plotted in Fig.~\ref{fig:PiGate} the time evolution of the local coherent energy ${E}\coh^{(A)}(t)$ for $t\in [0,T]$, the energetic contribution to entanglement $\hbar\omega_A C^2(t)$ computed from the state $\rho_{AB}(t)$, as well as the coherent-energy budget $\hbar \omega_A C^2(t) + E\coh^{(A)}(t)$. As expected for $\gamma = 0$, this budget is constantly equal to $\bar{E}\coh^{(A)}$, which appears as the conserved energetic resource along the process.  For $\gamma >0$, one observes that $\hbar \omega_A C^2(t) + E\coh^{(A)}(t) < \bar{E}\coh^{(A)}$. This is expected as the qubits evolve into a mixture. Moreover, the sum decreases over time: this can be seen seen as a continuous loss of coherent-energy budget, that gets locked into correlations with the environment. Further analyses based on models of Markovian environments would be necessary to confirm that trend. \\

\noindent {\em Three-qubit configuration.}---We finally consider a three-qubit protocol in which the state of qubit $A$ is mapped onto qubit $B$, then the state of qubit $B$ is mapped onto qubit $C$, through two successive LEP two-qubit interactions. The qubits are initially prepared in the product state $\ket{\Psi_{ABC}(t_0)} = \otimes_{m} \ket{\psi_m(t_0)}$ with $\ket{\psi_m(t_0)}= \sqrt{p}\ket{e_m} + \sqrt{1-p}\ket{g_m}$, $m=A,B,C$. This corresponds to a total coherent-energy budget $\bar{E}\coh = \sum_m \hbar \omega_m p(1-p)$. As above, we choose an equal excitation in each qubit to allow for complete coherence-to-entanglement conversions. Note that the considered LEP processes can be different from the phase gate considered above, hence such complete conversions do not necessarily require $p=1/2$. 

The first LEP takes place between $t_0$ and $t_1$ and prepares $\ket{\psi_{ABC}(t_1)} = \ket{\psi_{AB}(t_1)}\otimes \ket{\psi_C(t_0)}$, where $\ket{\psi_{AB}(t_1)} = \sqrt{p}\ket{e_A,e_B} + \sqrt{1-p}\ket{g_A,g_B}$. Thus at $t=t_1$, the coherent-energy budget allocated to $A$ and $B$ is entirely stored as bipartite entanglement, 
\beq \label{eq:GHZ1}
\bar{E}\coh^{(A)}+\bar{E}\coh^{(B)} = \hbar (\omega_A+\omega_B) C_{A|B}^2(t_1),
\eeq
where $C_{A|B}^2(t)$ denotes the scaled tangle of $\ket{\psi_{AB}(t)}$. Defining the non-local logical qubit $L$ by its eigenstates $\ket{i_{L}}:=\ket{i_A,i_B}$ ($i=g,e$) and transition frequency $\omega_{L} = (\omega_A+\omega_B)$ allows us to rewrite $\ket{\psi_{AB}(t_1)} \equiv \ket{\psi_{L}}=\sqrt{p}\ket{e_{L}}+\sqrt{1-p}\ket{g_{L}}$. Noticeably, the left-hand side of Eq.~\eqref{eq:GHZ1} can be rewritten as $\hbar \omega_L p(1-p)$, \ie~the coherent energy of the non-local qubit $L$ at time $t_1$. The first LEP interaction can thus be interpreted as a complete transfer of coherent energy from two local qubits at time $t_0$, $E\coh^{(A)}(t_0)+E\coh^{(B)}(t_0)$, into a single non-local one, $E\coh^{(L)}(t_1)$.

This perspective naturally extends to the second LEP interaction taking place between $t_1$ and $t_2$, that progressively entangles $C$ with the logical qubit $L$,  
\beq\label{eq:3qubit_process}
\ket{\Psi_{LC}(t)} = \sqrt{p}\ket{e_{L},M^e_C(t)} + \sqrt{1-p}\ket{g_{L},M^g_C(t)}, 
\eeq
with $t\in[t_1,t_2]$. $\ket{M_C^i(t)}$ denotes the state of qubit $C$ correlated with the state of the logical qubit $\ket{i_{L}}$, with $\ket{M_C^i(t_2)}=\ket{i_C}$. Note that the dynamics of $A$ and $B$ remains restricted to the Hilbert space spanned by $\{ \ket{g_L}; \ket{e_L} \}$ or {\it logical subspace}, allowing us to apply Eq.~\eqref{eq:Deficit-Coherence3} on $L$ at any time $t$ of the process:  \begin{equation}\label{eq:monogamy} 
\bar{E}\coh^{(L)} = E\coh^{(L)}(t) + \hbar\omega_{L} C^2_{L|C}(t). 
\end{equation} 
$C^2_{L|C}(t)$ quantifies the entanglement building up between $L$ and $C$, $E\coh^{(L)}(t)$ the coherent energy of $L$. Remarkably in the logical subspace, $E\coh^{(L)}(t)$ also equals $\hbar \omega_L C^2_{A|B}(t)$ (See EM) and as such, also tracks the entanglement still shared by $A$ and $B$. Thus, Eq.~\eqref{eq:monogamy} exposes a direct competition between two forms of entanglement. This energetic trade-off echoes the monogamy constraints of entanglement \cite{CofWoo00,Koashi04}. 

Applying Eq.~\eqref{eq:Deficit-Coherence3} to both $L$ and $C$ further shows that the second LEP interaction consumes the coherent energies $E\coh^{(L)}(t)$ and $E\coh^{(C)}(t)$ to generate higher-order correlations quantified by $\hbar(\omega_L+\omega_C)C_{L|C}^2(t)$, while preserving the total coherent-energy budget $\bar E\coh$. At $t=t_2$, the system reaches $\ket{\psi_{ABC}(t_2)}=\sqrt{1-p}\ket{g_A,g_B,g_C}+\sqrt{p}\ket{e_A,e_B,e_C}$, which corresponds to the state of a tripartite logical qubit of transition frequency $\omega_L+\omega_C$. Genuine tripartite correlations quantified by the total entanglement energy $\hbar(\omega_L+\omega_C)C^2_{L|C}(t_2)$ now store the entire coherent-energy budget $\bar{E}\coh$. Equivalently, this quantity can be interpreted as the coherent energy of the tripartite logical qubit. More generally, this protocol illustrates how successive LEP interactions promote coherence from local to increasingly non-local forms, converting local coherence into bipartite and subsequently multipartite entanglement while preserving the total coherent-energy budget.\\

\noindent {\em Conclusion.}---We introduced an energetic framework that identifies a fundamental constraint linking local coherence and qubit-qubit entanglement. For arbitrary pure two-qubit states, the coherent-energy deficit of each qubit is exactly proportional to the tangle, establishing a one-to-one correspondence between a locally measurable energetic quantity and an entanglement monotone. This correspondence yields a family of conservation laws for local-energy-preserving dynamics, where the coherent-energy budget can be reversibly redistributed between local coherence and entanglement. Our framework provides a direct energetic interpretation of local coherence-entanglement interconversion, leading to tight bounds on entanglement generation, explicit conversion efficiencies, and their degradation by mixedness. The controlled-phase gate illustrates how entanglement generation can be understood as the reversible transfer of local coherent energy into quantum correlations. Extending the analysis to three qubits reveals a hierarchy of coherence transfers, where coherence is progressively promoted from local degrees of freedom to bipartite and then multipartite forms of entanglement. Beyond the specific results obtained here, our results advance the emerging field of quantum energetics \cite{AuffevesElouard2026QuantumEnergetics}. It will be interesting to explore if and how entanglement generation under energy constraints can be optimized using these new metrics, in relation with former works that explicitly shown a quantitative link between work exchanges and coherent energy changes \cite{wenniger2023, monsel2020energetic, potts2025, Sam2024OBE}. Another interesting question would be to extend our framework to bipartite qudit systems and continuous variables, using for instance ergotropy \cite{Mukherjee16,Francica22,Adesso26}, or to different types of entanglement, like bound entanglement \cite{Horodecki98,HorodeckiRMP09}. \\

{\em Acknowledgments}---We would like to thank Travis J.~Baker and Alexssandre de Oliveira Junior 
for helpful discussions and comments on our manuscript.
This project is supported by the National Research Foundation, Singapore through the National Quantum Office, hosted in A*STAR, under its Centre for Quantum Technologies Funding Initiative (S24Q2d0009), and the Plan France 2030 through the projects NISQ2LSQ (Grant ANR-22-PETQ-0006), OQuLus (Grant ANR-22-PETQ-0013), and OECQ through BPI France.

\clearpage

\onecolumngrid

\begin{center}
    \textbf{\large End Matter}
\end{center}

\twocolumngrid
\appendix

\section{Comparison between ergotropy and coherent energy}

\begin{figure}[t]
 \centering
    \includegraphics[width=\linewidth]{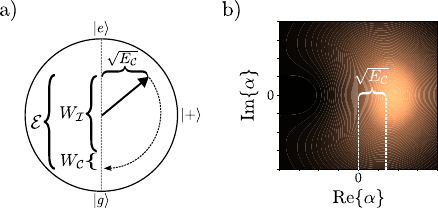}
    \caption{Geometric and phase-space representations of energetic quantities. We took $\hbar \omega =1$. a) Representation of the qubit state on the Bloch sphere. The coherent energy $E_{\mathcal{C}}$ equals the modulus square of the qubit's coherence, \ie, the radius of the Bloch vector. The unitary evolution it would undergo for the extraction of ergotropy is shown by the dotted line. The ergotropy $\mathcal{E}$, the incoherent work $W_{\mathcal{I}}$ and the coherent work $W_{\mathcal{C}}$ are shown in relation to this evolution. b) Q-representation of the bosonic field's state after the mapping of the qubit's state (See Fig.\ref{fig:Eg_Energetics}a and main text). The qubit's coherent energy equals the energy of the coherent component of the field.}
    \label{fig:Ergo_compare}
\end{figure}

The ergotropy ${\cal E}$ of a system is defined as the maximum energy one can extract from a quantum system via unitary operations. For a qubit $m$, it is related to the coherent deficit via
\beq 
{\cal E}^{(m)} = E_m - \frac{\hbar\omega_m}{2}\left(1 - \sqrt{1 - 4\frac{{\cal D}^{(m)}}{\hbar\omega_m}}\right).
\eeq
One can also subdivide the ergotropy into coherent $W^{(m)}\coh$ and incoherent contribution $W\incoh^{(m)}$ \cite{Shi2022Entanglement}, where the incoherent work is $W\incoh^{(m)} = 2E_m - \hbar\omega_m$ and the coherent work $W\coh^{(m)} = {\cal E}^{(m)} - W\incoh^{(m)}$. Fig.~\ref{fig:Ergo_compare}a displays a geometric interpretation of the ergotropy, coherent and incoherent work, making clear the distinction with the coherent and incoherent energies.



 \section{Work extraction in spontaneous emission processes}
In the main text, we described a process to extract the coherent energy by letting the qubit release its energy into a an initially empty bosonic mode (see Fig.~\ref{fig:Eg_Energetics}). The so-called spontaneous mode is defined by the annihilation operator $a_{\rm sp} = \int\dd t v_{\rm sp}(t) a(t)$, and the mode shape is given by $v(t) = \sqrt{\gamma\inv} e^{-\gamma(t - t_0)/2}$ for $t\geq t_0$ with $t_0$ being the start time of the interaction. The coherent energy of the qubit corresponds to the energy of the coherent component of the spontaneously emitted field. It can been seen in the Q-representation of the field, where the Husimi Q-function is defined as $Q(\alpha) = \bra{\alpha}\rho\ket{\alpha}/\pi$ \cite{WallsMilb94}. See Fig.~\ref{fig:Ergo_compare}b for an example Q-function of the field state after the emission.

\section{Mixed qubit-qubit state}
To gain further insights into the physical meaning of the coherent energy deficit in this general case, we express the joint density matrix as a mixture of joint pure states $\ket{\Psi_k}$, $\rho_{AB}= \sum_{k}q_k\op{\Psi_k}{\Psi_k}$, where $\sum_{k}q_k=1$. Note that the $\ket{\Psi_k}$ are not necessarily orthogonal. It is well known that this decomposition is not unique. We shall use the convenient operational view that it results from a preparation of the pure states $\ket{\Psi_k}$ with probability $q_k$, information on this preparation being accessible, or not. We show in \cite{SM} that the coherent energy deficit in each qubit can be expressed as a sum of two terms depending on the chosen decomposition,
\beq
{\cal D}^{(m)} = {\cal D}_{\rm Q}^{(m),\{k\}} + {\cal D}_{{\rm Cl}}^{(m),\{k\}}\,,
\eeq
where $\{k\}$ is a shorthand notation to refer to the mixture $\{ q_k;\ket{\Psi_k}\}$. The quantity ${\cal D}_{\rm Q}^{(m),\{k\}}$ reads \blk
\beq\label{eq:quant_deficit}
{\cal D}_{\rm Q}^{(m),\{k\}} = \hbar\omega_m{\mathbb E}_{k}\{C_k^2\}\,.
\eeq\blk
${\mathbb E}_{k}\{X\}$ denotes the expectation value of $X$ over $k$ and \blk $C_k = C[\ket{\Psi_k}\bra{\Psi_k}]$. ${\cal D}_{\rm Q}^{(m),\{k\}}$ \blk captures a deficit of quantum origin, due to the entanglement present in each pure state of the mixture. Conversely, the term ${\cal D}_{{\rm Cl}}^{(m),\{k\}}$ accounts for the loss of purity of the reduced state of qubit $m$, which does not stem from the entanglement with the other qubit,
\beq \label{eq:var}
{\cal D}_{{\rm Cl}}^{(m),\{k\}}= \hbar\omega_m{\mathbb V}_{k}\{\ex{\s{}^{(m)}}_k\} + (\hbar\omega_m)\inv{\mathbb V}_{k}\{E^{(m)}_k\} \,.
\eeq
Here, ${\mathbb V}_{k}\{X\} = {\mathbb E}_{k}\{|X|^2\} - |{\mathbb E}_{k}\{X\}|^2$ is the variance of $X$ over $k$, $\ex{\s{}^{(m)}}_k$ is the mean field of $\ket{\Psi_k}$, and $E_k^{(m)}$ its mean energy. 
${\cal D}_{{\rm Cl}}^{(m),\{k\}}$ is trivially positive. It vanishes when $\rho_{AB}$ is pure, or when all the pure states of the decomposition are characterized by the same local energies ($E_k = E_{k'}$) and mean fields ($\ex{\s{}^{(m)}}_k = \ex{\s{}^{(m)}}_{k'}$). Our energetic constraint becomes \blk
\beq \label{eq:Energy-cons-mix}
\bar{E}\coh^{(m)} = E\coh^{(m)} + \hbar\omega_m {\mathbb E}_{k}\{C_k^2\} +{\cal D}_{{\rm Cl}}^{(m),\{k\}}\,.
\eeq\blk
%
Applying the same reasoning as above, Eq.~\eqref{eq:Energy-cons-mix} shows that the energetic resource $\bar{E}\coh^{(m)}$ input in $\rho_{AB}$ can fuel qubit-qubit entanglement, here quantified by $\hbar\omega_m{\mathbb E}_{k}\{C_k^2\}$. Note that such an entanglement measure presupposes that information on the mixture is available. In this case, the efficiency of conversion defined for pure states extends to $\eta = \hbar(\omega_A + \omega_B){\mathbb E}_{k}\{C_k^2\}/ (\bar{E}\coh^{(A)} + \bar{E}\coh^{(B)})$. With respect to the pure case of same coherent energy $E\coh$ and potential coherent energy $\bar{E}\coh$, the efficiency is now lowered by the impurity of the joint state quantified by ${\cal L}^{\{k\}} ={\cal D}_{{\rm Cl}}^{(m),\{k\}} / (\bar{E}\coh^{(A)} + \bar{E}\coh^{(B)})$. Remarkably, the mixture minimizing this efficiency yields a direct equivalence to the scaled square concurrence of entanglement for the mixed joint state $\rho_{AB}$, namely \blk
\beq \label{eq:Neg-mix}
C^2[\rho_{AB}] = \min_{\{ k \}} {\mathbb E}_k\{C_k^2\}\,.
\eeq\blk
To see this, we use the qubit-qubit pure-state decomposition of Wootters that minimizes the \blk average concurrence \blk \cite{Wootters98}. This optimal decomposition entails at most four pure states of equal concurrence, which thus equals the concurrence of the mixed state. \blk From this, by using Jensen's inequality ${\mathbb E}_k\{C_k^2\}\geq ({\mathbb E}_k\{C_k\})^2$ and realizing that $x^2$ is a monotonic function for $x\geq 0$, one obtains Eq.~\eqref{eq:Neg-mix}. Note, unlike the pure state case, $C^2[\rho_{AB}]$ is not equivalent to the squared negativity \cite{Miranowicz04PRA}.

\section{Three-qubit Coherent Energy}
In the process we are considering, the logical qubit progressively becomes entangled with the third qubit via \erf{eq:3qubit_process}. Initially, at $t_1$, the states of qubit $C$ are indistinguishable, \ie, $\ket{M_C^g(t_1)} = \ket{M_C^e(t_1)}$, meaning that the $AB|C$ partition is separable. Furthermore, the coherent energy in the qubit $L = AB$ is given by the entanglement between qubits $A$ and $B$, \ie, $E\coh^{(L)}(t_1) = \hbar(\omega_A + \omega_B)C_{A|B}^2(t_1)$, by construction. As the entangling process between qubit $L$ and $C$ begins, the distinguishability, $\epsilon(t) = \ip{M_C^g(t)}{M_C^e(t)}$, of qubit $C$'s states begins to decrease, eventually dropping to zero. Computing the coherent energy in qubit $L$ during this process, one simply needs to compute the reduced state $\rho_L \equiv \rho_{AB}$, which gives
\beq
\begin{split}
\rho_{AB}(t) &= p\op{ee}{ee} + \epsilon(t)\sqrt{(1 - p)p}\op{ee}{gg} +\\& \epsilon^*(t)\sqrt{(1 - p)p}\op{gg}{ee} + (1-p)\op{gg}{gg}\,.
\end{split}
\eeq
Computing the coherent energy for qubit $L = AB$, one finds $E\coh^{(L)} = \hbar\omega_L|\ex{\s{L}}|^2 = \hbar\omega_L|\epsilon(t)|^2 (1-p)p$, where $\s{L} = \op{g_L}{e_L}$. This, however, is proportional to the entanglement (quantified by the tangle) between qubits $A|B$. 

To relate the coherent energy of qubit $L$ to the entanglement between $A$ and $B$, we remind the reader that the concurrence is given, for mixed states, by ${\mathbb C}_{A|B}[\rho] = \max(0, \lambda_0 - \lambda_1 - \lambda_2 - \lambda_3)$ where $\lambda_i$ denoting the eigenvalues in decreasing order of $R = \sqrt{\sqrt{\rho}\tilde\rho\sqrt{\rho}}$ matrix with $\tilde{\rho} = (\s{y}\otimes\s{y})\rho_{AB}\tp (\s{y}\otimes\s{y})$. As we are working in a reduced subspace, it is easy to see that two of the eigenvalues of $R$ will be zero. Computing the remaining eigenvalues, one finds $\lambda_0 = \sqrt{(1-p)p} (1 + |\epsilon(t)|)$ and $\lambda_1 = \sqrt{(1-p)p} (1 - |\epsilon(t)|)$, giving ${\mathbb C}[\rho_AB(t)] = 2|\epsilon(t)| \sqrt{(1-p)p}$. Remembering that $\omega_A + \omega_B = \omega_L$ and $C_{A|B} = {\mathbb C}_{A|B}/2$, we get that the total entanglement energy between qubits $A$ and $B$, $E_{\rm ent}(t) = \hbar(\omega_A + \omega_B) C^2_{A|B}(t)$, is equal to the coherent energy of the logical qubit $L$ at all times during the process, \ie, $E\coh^{(L)}(t) = \hbar\omega_L C^2_{A|B}(t)$. Note, it should be stated that this is only the case for the specific process that we are considering. More generally, one may have to take into account classical correlations formed during the process, requiring the use of the mixed state case presented previously.

\onecolumngrid

\begin{center}
    \textbf{\large Supplementary Material: An Energetic Constraint for Qubit-Qubit Entanglement}
\end{center}

\renewcommand{\thefigure}{S\arabic{figure}}
\setcounter{figure}{0}

\renewcommand{\theequation}{S.\arabic{equation}}
\setcounter{equation}{0}

\renewcommand{\thesubsection}{S.\arabic{subsection}}
\setcounter{subsection}{0}

\subsection{Computation of the coherent deficit for each mode}
Let us begin by considering the coherent energy deficit for mode $A$. One can write the general two-qubit state, $\ket{\Psi_{AB}} = p_{00}\ket{0_A,0_B} + p_{01}e^{-i\phi_{01}}\ket{0_A,1_B} + p_{10}e^{-i\phi_{01}}\ket{1_A,0_B} + p_{11}e^{-i\phi_{11}}\ket{1_A,1_B}$, as $\ket{\Psi_{AB}} = \sqrt{1 - p_A} \ket{0,M^{A}_{0}} + \sqrt{p_A} \ket{1,M^{A}_{1}}$, where 
\begin{align}
\ket{M^{A}_{0}} &= \frac{1}{\sqrt{1 - p_A}}\left(\sqrt{p_{00}}\ket{0} + \sqrt{p_{01}}e^{-i\phi_{01}}\ket{1}\right)\,,\\
\ket{M^{A}_{1}} &= \frac{1}{\sqrt{p_A}}\left(\sqrt{p_{10}}e^{-i\phi_{10}}\ket{0} + \sqrt{p_{11}}e^{-i\phi_{11}}\ket{1}\right)\,.
\end{align}
and $p_A = p_{10} + p_{11}$. The coherent energy deficit for mode $A$ is given by ${\cal D}^A = (1 - |\epsilon_A|^2)\bar{E}_{\cal C}^A$, where $\epsilon_A = \ip{M^{A}_{0}}{M^{A}_{1}}$ and the nominal coherent energy is $\bar{E}_{\cal C}^A = \hbar\omega_A p_A (1 - p_A)$. Computing the distinguishability ($|\epsilon_A|^2$), one obtains
\beq\label{eq:epA}
|\epsilon_A|^2 = \frac{1}{p_A(1 - p_A)}\left(p_{00}p_{10} + p_{01}p_{11} + 2\sqrt{p_{00}p_{10}p_{01}p_{11}}\cos{\Delta \phi}\right)\,,
\eeq
where $\Delta \phi = \phi_{11} - \phi_{10} - \phi_{01}$.
Substituting this into the coherent energy deficit along with the nominal coherent energy,
\beq
\bar{E}_{\cal C}^A = \hbar\omega_A p_A (1 - p_A) = \hbar\omega_A (p_{00} + p_{01})(p_{10} + p_{11})\,,
\eeq
one finds that 
\beq\label{eq:wp_Engy_A}
{\cal D}^A = \hbar\omega_A \left(p_{01}p_{10} + p_{00}p_{11} - 2\sqrt{p_{00} p_{01} p_{10} p_{11}}\cos{\Delta \phi}\right)\,.
\eeq

To compute the coherent energy deficit for subsystem $B$, we rewrite the joint system state with subsystem $A$ acting as the meter state, \ie, $\ket{\Psi_{AB}} = \sqrt{1 - p_B} \ket{M^{B}_{0},0} + \sqrt{p_B} \ket{M^{B}_{1},1}$, where $p_B = p_{01} + p_{11}$ and 
\begin{align}
\ket{M^{B}_{0}} &= \frac{1}{\sqrt{1 - p_B}}\left(\sqrt{p_{00}}\ket{0} + \sqrt{p_{10}}e^{-i \phi_{10}}\ket{1}\right)\,,\\
\ket{M^{B}_{1}} &= \frac{1}{\sqrt{p_B}}\left(\sqrt{p_{01}}e^{-i\phi_{01}}\ket{0} + \sqrt{p_{11}}e^{-i \phi_{11}}\ket{1}\right)\,.
\end{align}
Following a similar analysis to that for subsystem $A$, one finds that the coherent energy deficit for subsystem $B$ is given by ${\cal D}^B = (1 - |\epsilon_B|^2)\bar{E}_{\cal C}^B$, where $\epsilon_B = \ip{M^{B}_{0}}{M^{B}_{1}}$ and the nominal coherent energy is $\bar{E}_{\cal C}^B = \hbar\omega_B p_B (1 - p_B)$.
Computing the distinguishability ($|\epsilon_B|^2$), we get
\beq\label{eq:epB}
|\epsilon_B|^2 = \frac{1}{E^B(1 - E^B)}\left(p_{00}p_{01} + p_{10}p_{11} + 2\sqrt{p_{00}p_{10}p_{01}p_{11}}\cos{\Delta \phi}\right)\,.
\eeq
Substituting this into the coherent energy deficit along with the nominal coherent energy,
\beq
\bar{E}_{\cal C}^B = \hbar\omega_B p_B (1 - p_B) = \hbar\omega_B(p_{00} + p_{10})(p_{01} + p_{11})\,,
\eeq
we obtain
\beq
{\cal D}^B = \hbar\omega_B\left(p_{01}p_{10} + p_{00}p_{11} - 2\sqrt{p_{00} p_{01} p_{10} p_{11}}\cos{\Delta \phi}\right)\,.
\eeq
We can see that this expression is proportional to the coherent energy deficit for subsystem $A$ in \erf{eq:wp_Engy_A} and thus we have ${\cal D}^A/\hbar\omega_A = {\cal D}^B/\hbar\omega_B$.

\subsection{Maximizing the coherent defict under fixed energy constraints}
Here we address the problem of maximizing the coherent deficit,
\beq
(\hbar\omega_m)\inv {\cal D}^m = p_{01}p_{10} + p_{00}p_{11} - 2\sqrt{p_{00} p_{01} p_{10} p_{11}}\cos{\Delta \phi}\,.
\eeq
under a fixed energy constraint for $A$ and $B$. That is, given $E^A = \hbar\omega_A p_A$ and $E^B = \hbar\omega_B p_B$, what is the maximum coherent deficit possible? We can first see that, since the energy constraints, $p_A = p_{10} + p_{11}$ and $p_B = p_{01} + p_{11}$, place no constraints on the relative phases in the state, we can always find a state such that $\cos{\Delta\phi} = -1$. Utilizing the energy constraints, as well as the normalization constraint $p_{00} + p_{10} + p_{01} + p_{11} = 1$, the coherent deficit becomes
\beq\label{deficit_2}
(\hbar\omega_m)\inv{\cal D}(p_{11}) = \left(\sqrt{(p_A - p_{11})(p_B - p_{11})} + \sqrt{(1 + p_{11} - p_A - p_B)p_{11}}\right)^2\,,
\eeq
where $p_{11}\in[0, \min\{p_A,p_B\}]$. By the definition of the coherent deficit, we know that it must be bounded by $\min\{\bar{E}\coh^A,\bar{E}\coh^B\}$. It is easy to show by substitution into \erf{deficit_2}, that choosing
\beq
p_{11} = E_A\left|\frac{1 - p_A - p_B}{1 - 2p_A}\right|\,, \qquad {\rm or} \qquad p_{11} = p_B\left|\frac{1 - p_A - p_B}{1 - 2p_B}\right|\,,
\eeq
leads to ${\cal D} = \bar{E}\coh^A$ or $\bar{E}\coh^B$, respectively. Let us consider the first case. Specifically, it is easy to see that whenever 
$E_A\leq E_B \leq \half$, and hence $\bar{E}\coh^A\leq\bar{E}\coh^B$, the term in the absolute value is less than unity, meaning that this $p_{11} \in[0,E_A]$ 
is a valid assignment, and the maximum coherent deficit is ${\cal D}_{\max} = \bar{E}\coh^A$. Similarly, for the second assignment, we can see that it is
always valid when $E_B\leq E_A \leq \half$, and hence $\bar{E}\coh^B\leq\bar{E}\coh^A$, and ${\cal D}_{\max} = \bar{E}\coh^B$. The other regimes are 
simply obtained by simple symmetry arguments. Thus we have that ${\cal D}_{\max} = C^2_{\max} = \min\{\bar{E}\coh^A,\bar{E}\coh^B\}$.

\subsection{The energies of maximally correlated pure states}
Here we show that, under the optimal conversion, $|\epsilon_A| = |\epsilon_B| = 0$, the energies of qubit $A$ and qubit $B$ must be perfectly correlated/anti-correlated ($p_A = p_B$ or $1 - p_B$). In fact, we prove something slightly stronger, that being, $|\epsilon_A| = |\epsilon_B|$ iff the energies are perfectly correlated/anti-correlated. 

Let us begin with the forward implication. If $|\epsilon_A| = |\epsilon_B|$, then it is easy to show from \erf{eq:epA} and \erf{eq:epB} that $(p_{00} - p_{11})(p_{10}-p_{01}) = 0$. Thus, either $p_{00} = p_{11}$ or $p_{10} = p_{01}$. The latter case leads to the perfect correlation scenario, since $p_A = p_{11} + p_{10} =  p_{11} + p_{01} = p_B$. For the former, by normalization of the state, we have that 
\beq
1 = p_{00} + p_{01} + p_{10} + p_{11} = p_{01} + p_{10} + 2p_{11} = p_A + p_B\,,
\eeq
yielding $p_A = 1 - p_B$. 

The reverse case is trivial to prove since ${\cal D}_A/\hbar\omega_A = {\cal D}_B/\hbar\omega_B$ and the fact that the nominal coherent energies are proportional in both the perfectly correlated and anti-correlated cases.

\subsection{Derivation of Eq. (7-9)}
Let us begin with by using the definition of the coherent energy deficit for subsystem $m$,
\beq
{\cal D}^{m} = \bar{E}\coh^{m} - E\coh^{m}\,.
\eeq
Rewriting the nominal coherent energy as $\bar{E}\coh^{m} = E^{m} - \bar{E}\incoh^m$, we have
\beq\label{seq:Correlation_Energy-2}
{\cal D}^{m} = E^m - \bar{E}\incoh^{m} - E\coh^{m}\,.
\eeq
Now, for the pure state decomposition $\rho = \sum_k q_k \op{\Psi_k}{\Psi_k}$, the energy of subsystem $m$ is $E^m = \sum_{k}q_k E^{m,k}$. Furthermore, for each pure state $\ket{\Psi_k}$, we can decompose the total energy as
\beq
E^{m,k} = E\coh^{m,k} + \hbar\omega_m C_{k}^2 + \bar{E}\incoh^{m,k}\,,
\eeq
where $C_k$ is the (scaled) concurrence of the state $\ket{\Psi_k}$.
Substituting this into \erf{seq:Correlation_Energy-2}, one obtains
\beq
{\cal D}^{m} = \sum_{k} q_k \hbar\omega_m C_k^{2} + \left(q_k E\coh^{m,k} - E\coh^{m}\right) + \left(q_k \bar{E}\incoh^{m,k} - \bar{E}\incoh^{m}\right)\,.
\eeq

From this point, we define ${\cal D}_{\rm Q}^{m,\{k\}} = \hbar\omega_m \sum_{k} q_k C_k^{2}$ and ${\cal D}_{\rm Cl}^{m,\{k\}} = \sum_{k}\left(q_k E\coh^{m,k} - E\coh^{m}\right) + \left(q_k \bar{E}\incoh^{m,k} - \bar{E}\incoh^{m}\right)$, leading us to the energetic splitting of the coherent deficit into ${\cal D}^{m} = {\cal D}_{\rm Q}^{m,\{k\}} + {\cal D}_{\rm Cl}^{m,\{k\}}$.

\subsection{Energetic-Informational understanding}
In a footnote in the main text (Ref.~[12]), we stated that the name ``coherent energy'' could also be motivated from a resource theoretic perspective. In particular, we stated that the coherent energy was related to a measure of coherence in the energy basis. Here we elaborate. In the resource theory of coherence, one way to quantify how coherent a state is in a fixed basis it to measure how far the state is from the set of incoherent states in said basis. In our case, the fixed basis is the energy basis and the set of incoherent states are ${\cal I} = \{\rho = q\op{1}{1} + (1 - q)\op{0}{0}\}_q$. As it turns out, for any of the distance measures based on the $l_p$ matrix norms or the Schatten $p$-norms \cite{StrPle17}, the minimum distance to the incoherent set, up to a constant prefactor which is irrelevant, is
\begin{align}
{\cal C}[\rho] = k \min_{\sigma\in{\cal I}} ||\rho - \sigma|| =  |\ex{a}|
\end{align}
where $||A||$ is a simplified notation for any of the aforementioned matrix norms. Thus, $E^m_{\cal C} = \hbar\omega_m {\cal C}^2$ is the energy that arises purely due to the coherence in the energy basis, \ie, the {\em coherent energy}.

This connection enables us to view the coherent energy deficit from a more information theoretic perspective. In particular, in the pure bipartite qubit case, for qubit $m$, we have ${\cal D}^m = \bar{E}_{\cal C} - \hbar\omega_m({\cal C}^m)^2 = \hbar\omega_m C^2$, where we have used $E_{\cal C} = \hbar\omega_m{\cal C}^2$. We can also recognize that the nominal coherent energy, for qubits, is equivalent to the uncertainty in the energy operator $(\hbar\omega_m)\inv(\Delta E_m)^2 = \hbar\omega_m (\ex{(\s{m}\dg \s{m})^2} - \ex{\s{m}\dg \s{m}}^2) = \hbar\omega_m p_m(1 - p_m)$. This allows us to rewrite the coherent energy deficit as
\beq
(\hbar\omega_m)^{-2}\left(\Delta E_m\right)^2 = \left({\cal C}^m\right)^2 + C^2.
\eeq
Here, we can rather view the trade-off between coherence and entanglement as being due to the energy uncertainty of the system. Furthermore, in this form it is easier to see why, in the mixed state case, additional energy uncertainty due to the mixedness appears in the coherent deficit, with the uncertainty in the coherence appearing similarly. 

As an aside, this form may potentially provide a simple/alternative route to defining the nominal coherent energy for qudit systems, as the energy uncertainty (with an appropriate unit normalization).

\end{document}